**RESEARCH ARTICLE**

**Journal of COMPUTATIONAL CHEMISTRY** **WILEY**

# On the sensitivity of computed partial charges toward basis set and (exchange-)correlation treatment

## Nisha Mehta [ORCID] | Jan M. L. Martin [ORCID]

Department of Molecular Chemistry and
Materials Science, Weizmann Institute of
Science, Rehovot, Israel

**Correspondence**
Jan M. L. Martin, Department of Molecular
Chemistry and Materials Science, Weizmann
Institute of Science, 7610001 Rehovot, Israel.
Email: gershom@weizmann.ac.il

**Present address**
Nisha Mehta, School of Chemistry,
The University of Melbourne, Melbourne,
Victoria, Australia.

**Abstract**

Partial charges are a central concept in general chemistry and chemical biology, yet dozens of different computational definitions exist. In prior work [Cho et al., *ChemPhysChem* **21**, 688-696 (2020)], we showed that these can be reduced to at most three 'principal components of ionicity'. The present study addressed the dependence of computed partial charges $q$ on 1-particle basis set and (for WFT methods) $n$-particle correlation treatment or (for DFT methods) exchange-correlation functional, for several representative partial charge definitions such as QTAIM, Hirshfeld, Hirshfeld-I, HLY (electrostatic), NPA, and GAPT. Our findings show that semi-empirical double hybrids can closely approach the CCSD(T) 'gold standard' for this property. In fact, owing to an error compensation in MP2, CCSD partial charges are further away from CCSD(T) than is MP2. The nonlocal correlation is important, especially when there is a substantial amount of nonlocal exchange. Employing range separation proves to be "mostly" not advantageous, while global hybrids perform optimally for 20%–30% Hartree-Fock exchange across all charge types. Basis set convergence analysis shows that an augmented triple-zeta heavy-aug-cc-pV(T+d)Z basis set or a partially augmented jun-cc-pV(T+d)Z basis set is sufficient for Hirshfeld, Hirshfeld-I, HLY, and GAPT charges. In contrast, QTAIM and NPA display slower basis set convergence. It is noteworthy that for both NPA and QTAIM, HF exhibits markedly slower basis set convergence than the correlation components of MP2 and CCSD. Triples corrections in CCSD(T), denoted as CCSD(T)-CCSD, exhibit even faster basis set convergence.

## 1 | INTRODUCTION

Partial charges are a central concept familiar to every chemist and most chemical biologists. Yet the concept does not correspond to any single quantum mechanical observable, and dozens of competing computational definitions are in existence. (See, e.g., References 1–7 for reviews.)

Parr et al.[8] went as far as to refer to partial charges as *noumena*, that is, purely intellectual constructs. (The term *noumenon* was originally coined by Immanuel Kant in the context of his philosophy: the name derives from Greek *noös*, knowledge or cognition, as opposed to a phenomenon—which is observable and measurable, from Greek *phainein*='to show'.)

In response, Bader and Matta retorted[9,10] that there is one definition that *can* be defined as the expectation value of a quantum mechanical operator: QTAIM (quantum theory of atoms in molecules), also popularly known as AIM or 'Bader charges'.[11] These can be obtained by integrating the electron density over an operator which consists of a Heaviside function that is unity inside a certain set of zero-flux surfaces and zero outside. Furthermore, they argue that the idea of partial charges being a mere mental construct flies in the face of long practical chemical experience.

In Cho et al.(CSESEM),[12] one of us (JMLM) argued for a way out of this conundrum: to consider partial charges 'proxy variables' for a deeper concept that eludes direct measurement, in this case, the "[molecular] ionicity" conjectured by Meister and Schwarz.[13] (Examples of proxy variables in wide usage are GDP per capita and PPP as proxies for economic welfare, or standardized test scores as proxies for general intelligence or scholastic aptitude. Tree ring measurements as proxies for past temperatures are an example from paleoclimatology. Useful as proxies can be, one must never lose sight of the distinction between a proxy and a measurement.)

CSESEM considered, at one single level of electronic structure theory (namely, PBE0[14,15] in the def2-TZVPP basis set[16]) some two dozen different charge distribution definitions for a very large (over 2000 molecules, over 30,000 individual partial charges) dataset of main-group molecules (specifically, the closed-shell part of GMTKN55[17,18] excluding species with trivially-zero partial charges, such as homonuclear diatomics). We then subjected the partial charges dataset to principal component analysis, and found that nearly all the variation in the dataset could be described by two or, at most, three principal components (PCs): PC1 has all definitions going in sync and corresponds to the Meister-Schwarz "ionicity", while PC2 pits QTAIM (and to a lesser extent, Cioslowski's GAPT or Generalized Atomic Polar Tensor[19]) against all the others (Richter et al.[20] proposed a possible rationalization for the similarity in behavior of QTAIM and GAPT). Finally, PC3 mainly pits electrostatic charges against their orbital-based counterparts.

Upon a subsequent variable reduction analysis,[21] CSESEM concluded that the most compact description of the dataset is by three variables: QTAIM; HLY (Hu-Lu-Yang[22]) or some other electrostatic potential (ESP)-based charge such as CHELPG[23] or Merz-Singh-Kollman;[24] and natural population analysis (NPA)[25] or some other orbital-based charge. Intriguingly, these three partial charge types correspond to one representative from each of the three main charge classes in the 'taxonomy' of Corminboeuf and coworkers:[26] 1. fitting to an observable quantity like the ESP; 2. partitioning in terms of atomic orbitals; 3. partitioning in terms of the electron density.

Cramer and Truhlar[1] have put forward a different four-way taxonomy, further subdivided later by CSESEM. We shall recapitulate it below:

1. Class I charges are directly obtained from experiment, for example, from observed deformation densities[27,28] or the atomic electronegativities through the electronegativity equalization principle.[29] A reviewer commented that atomic polar tensor charges[19] belong in Class I since they can be extracted experimentally from a combination of geometries, harmonic frequencies, dipole moments, and infrared intensities: see the original APT papers[30,31] for more details, as well as Reference 32 for an example covering 167 atoms in 67 diverse molecules. While this is arguably true for APT in the same way as it would be for Hirshfeld-type charges (normally Class IIb1) if obtained from experimental deformation densities, this classification seems less appropriate for GAPT.

2. Class IIa is obtained from partitioning orbitals:
   a. Class IIa1 consists of the original Mulliken population analysis[33] and later variants, for example, Bickelhaupt,[34] when applied *in the whole basis set*. All of these exhibit pathological sensitivity to said basis set, and therefore do not satisfy even the weakened Cioslowski-Surjan[35] observability criterion.
   b. The latter is satisfied by Class IIa1M, in which the MOs are first projected onto a minimal basis set, and the methods from IIa1 are then subsequently applied to the projection: examples are MBS-Mulliken[36] and MBS-Bickelhaupt.[12]
   c. Class IIa2 charges are based on some form of natural or intrinsic atomic orbital, such as NBO[25,37] of Weinhold and coworkers, and IAOs (intrinsic atomic orbitals,[38] which are functionally equivalent[39] to Ruedenberg's quasi-atomic orbitals[40])

3. Class IIb is obtained by partitioning the electron density:
   a. Class IIb1 corresponds to 'fuzzy' (i.e., non-disjoint) domains, where the density at a given point in space can contribute to more than one atom, such as in the original Hirshfeld population analysis[41,42] and its later iterative variants.[43–46]
   b. Class IIb2 corresponds to disjoint domains, where space is partitioned into nonoverlapping 'cells'. QTAIM, as explained above, is the paradigmatic example; another is VDD or Voronoi Deformation Density.[47]

4. Class III are obtained by fitting to some physical observable extracted from the wave function *viz.* electron density
   a. Class IIIa are fitted to the electrostatic potential, such as CHELP,[48] CHELPG,[23] Merz-Singh-Kollman,[24] and HLY[22] charges. All of these fit the electrostatic potential obtained from a network of point partial charges to the ESP obtained from the WFT- or DFT-computed electron density; they just differ in the sampling procedures for the grid.
   b. Class IIIb are obtained from other electric properties, such as, for example for GAPT,[19] one-third the atomic trace of the dipole moment derivative matrix.

5. Finally, Class IV was introduced by Cramer and Truhlar,[1] and involves a semiempirical adjustment of some well-defined Class II or III charge for better reproduction of one or more physical observables (e.g., the dipole moment in Charge Model Five, CM5[49]).

The main question left unanswered by CSESEM was the sensitivity of the individual partial charges to basis set and electron correlation treatment (for WFT methods) or exchange-correlation functional (for DFT methods). This question was only superficially touched on in an older study on a small sample by De Proft, Martin, and Geerlings.[50] We will address it in depth in the present contribution. To remove all doubt, the present paper does not concern itself with the question as to which is the 'better' or more chemically meaningful partial charge

model—the answer there to depends on which aspect of 'chemical ionicity' is of interest, as discussed in CSESEM as well as in, for example, Reference [51]—but rather with the question, for a given partial charge type, how to obtain converged values in terms of 1-particle basis set and electron correlation treatment.

## 2 | COMPUTATIONAL DETAILS

As evident from Reference [12], in order to ensure a holistic view of the provided database, we need to consider at the very least one representative from each of the three classes defined by Corminboeuf and co-workers.[26] (Classes in the extended Cramer-Truhlar classification, *vide supra*, are given in parentheses.)

1. Population analysis based on the electron density: obviously QTAIM[11] (Class IIb2), but we also have included Hirshfeld-I[43] as well as the original/'vanilla' Hirshfeld[41] (both Class IIb1).
2. Population analysis based on the atomic orbitals: Natural Population Analysis (NPA, Class IIa2),[25,37] as obtained from the NBO7 program.[52]
3. Population analysis based on the electrostatic potential: we semi-arbitrarily selected Hu-Lu-Yang[22] (HLY, Class IIIa) charges as their implementation in Gaussian 16[53] is numerically somewhat more stable than the others, but any of the other electrostatic charge variants (CHELPG,[23] Merz-Singh-Kollman[24]) would have led to essentially the same results (see p. 691, left-hand column, of Reference [12]).
4. In addition, we also explored partial charges based on the trace of the dipole moment derivatives, that is, GAPT[19] (generalized atomic polarizability tensor, class IIIb) charges. These can be related to an observable, namely the infrared intensity. (For more discussion, see Bruns and coworkers,[20] as well as the older References [30,31].)

The convergence with respect to basis set and electron correlation method of CM5[49] (Charge Model Five, the paradigmatic example of Class IV) is, by construction, *identical* to that of 'vanilla' Hirshfeld and hence will not be discussed separately here. Readers interested in the CM5 charges themselves, on account of their superior performance in mimicking molecular dipole moments, can find them in the electronic supporting information (ESI).

We will refrain from discussing the most popular charge type, Mulliken population analysis, as it does not satisfy even the Cioslowski-Surjan criterion:[35] Mulliken charges do not converge to a stable limit as the 1-particle basis set and n-particle correlation treatment approach completeness, and are in fact well known to behave erratically.[7,35] Perhaps the most lucid explanation why this happens is given by Frank Jensen on p. 319 of Reference [7]:

> A reasonable description of the wave function can be obtained by a Hartree-Fock single determinant with a DZP basis set. An equally good wave function (in terms of energy) may be constructed by having a very large number of basis functions centered on oxygen, and none on the hydrogens (a DZP quality basis set on both oxygen and hydrogen gives an energy similar to having a 5ZP quality basis set on oxygen only). The latter will, according to the above population analysis, have a +1 charge on hydrogen and a −2 charge on oxygen. Worse, another equally good wave function may be constructed by having a large number of basis functions only on the hydrogens. This will give charges of −4 for each of the hydrogens and +8 for the oxygen. Alternatively, the basis functions can be taken to be non-nuclear-centered, in which case the electrons are not associated with any nuclei at all [leading to 'atomic'] charges of +1 and +8!

Carrying out reference calculations at levels of theory that assure convergence would be impossible for the entire GMTKN55[18] dataset. In order to have a representative sample of first-and second-row molecules that is still somewhat computationally tractable, we selected the closed-shell molecules from the W4-17 computational thermochemistry benchmark[54]—an expanded version of W4-11[55] which is one of the constituent parts of GMTKN55. This set covers a variety of inorganic and organic species with different bonding situations (including pseudo-hypervalent species) and bond orders, as well as a broad range of static correlation degrees ranging from purely dynamical (e.g., n-butane) to strongly multireference (e.g., ozone) Upon excluding species with trivial charge distributions (such as homonuclear diatomics and tetrahedral $P_4$), we were left with 153 unique species. The original geometries in the W4-11 (and hence GMTKN55) dataset, which had been optimized at the CCSD(T)/cc-pV(Q+d)Z level, were used as is in all calculations and not optimized further. Furthermore, we omitted BN from our analysis due to its well-known pathological multireference character causing erratic results.

All electronic structure calculations were carried out using either Gaussian 16[53] or MOLPRO 2022.3.[56] running on the Faculty of Chemistry HPC facility 'ChemFarm' at the Weizmann Institute. Post-processing to extract certain charge types was carried out using NBO7,[52] AIMALL[57] for the QTAIM charges, and Multiwfn[58] for some others.

Gaussian was utilized for Hirshfeld, Hirshfeld-I, and HLY charges at the DFT (including double hybrid), HF, MP2, and CCSD levels, as well as for GAPT charges (which, as they require the evaluation of dipole moment derivatives, are obtained as by-products of analytical frequency calculations) at the DFT (including double hybrid), HF, and MP2 levels. All DFT calculations using Gaussian employed the `int (grid = ultrafine)` integration grid.

For some species, we encountered erratic HLY charges, which did not follow any obvious pattern across molecules or exchange-correlation functionals. We resolved this issue by increasing the ESP sampling grid size using `IOp(6/60)`. Somewhat to our surprise, we discovered that this option also affects the integration grid option used in Hirshfeld (and hence CM5) and Hirshfeld-I charges; hence we set `IOp(6/60 = 5)` throughout to ensure the grid is sufficiently fine for both ESP and Hirshfeld-derived charges.

MOLPRO 2022.3[56] was employed for GAPT charges at the CCSD and CCSD(T) level, as well as NPA charges at the CCSD(T) level. The CCSD(T) NBO charges were obtained from the first-order reduced density matrix through MOLPRO's NBO7 interface. At first, the GAPT charges were evaluated by numerically differentiating analytical CCSD(T) dipole moments with respect to nuclear coordinates; operationally, they were obtained as by-products of semi-numerical vibrational frequency calculations. However, this approach becomes costly and cumbersome beyond triatomics or at most tetratomics; hence, for larger molecules, we instead carried out finite differentiation of analytical geometric gradients with respect to external fields of $\pm 0.005$ a.u. This requires at most six gradient evaluations (absent symmetry); for the smaller molecules, we verified that the two different procedures yielded the same results to four decimal places. In this manner, we were able to evaluate GAPT charges for nearly all of W4-17 at the CCSD(T)/haVTZ+d level and a large subset at the haVQZ+d level.

NPA charges were extracted using the NBO7 program, which was interfaced with Gaussian and MOLPRO. We utilized the AIMALL[57] program to compute QTAIM charges. The AIMALL calculations employed the `bim=auto` Bayesian integration grid; our tests using the `bim=promega5` option revealed that we already obtained converged results with the `bim=auto` grid size.

A list of DFT and WFT methods evaluated in this article can be found in the first column of Table 1. There are literally hundreds of DFT exchange-correlation functionals nowadays (see, e.g., the many entries in the LibXC library[59]), and it would be impossible to present an exhaustive survey here. Hence, we have merely selected some representative examples on each rung of Jacob's Ladder.[60] In addition, in order to single out the behavior of specific components, we have (for example) considered not only PBE, but also PBE exchange alone, as well as PBE exchange with PW91LDA correlation (i.e., omitting the semilocal correlation correction); in the double hybrids, we for example, considered B2GP-PLYP as well as the same without the PT2 term, and B2GP exchange without correlation. (It should be noted that the parameters employed in the revDSD-PBEP86[61] double hybrid correspond to the revDSD-PBEP86-D3(BJ) variant. Obviously, the D3(BJ) empirical dispersion correction[62] will not contribute directly to the partial charges; if the geometry were optimized at the same level, it might contribute indirectly to the partial charges, particularly for noncovalent interactions.)

The basis sets considered are the correlation-consistent basis sets developed by Dunning and co-corkers:[63–66] we utilized cc-pVnZ for hydrogen, aug-cc-pVnZ for first-row elements, and aug-cc-pV(n+d)Z basis sets for second-row p-block elements, with $n = D, T, Q$ and 5. Throughout this article, this basis set combination is abbreviated as "haVnZ+d"; in the 'calendar' notation,[67,68] it corresponds to jul-cc-pV(n+d)Z. (We note in passing that, as was already shown by one of us,[69] that for the central second-row atom in high oxidation states X one has in species like $SO_3$ or $HClO_4$, inclusion versus omission of the additional high-exponent $d$ function can increase versus decrease the NPA/NBO partial charges on X by as much as 0.3, as the 3d orbitals in high oxidation states sink low enough to be able to act as a back-donation recipient from chalcogens and halogens.) In addition to that, we also briefly considered the Weigend-Ahlrichs' "def2" basis

**TABLE 1** Summary of root mean square deviations (RMSDs) for NBO7 and GAPT partial charges for closed-shell molecules in the W4-17 thermochemistry benchmark.

| | haVTZ+d | | haVQZ+d | |
|---|---|---|---|---|
| | NPA | GAPT | NBO7 | GAPT |
| SVWN5[73] | 0.050 | 0.042 | 0.049 | 0.044 |
| PBE$_x$ | 0.014 | 0.032 | 0.014 | 0.035 |
| PBE[74] | 0.030 | 0.032 | 0.030 | 0.034 |
| PBE-PW91LDA[75] | 0.012 | 0.027 | 0.012 | 0.030 |
| B88$_x$ | 0.012 | 0.033 | 0.012 | 0.036 |
| BLYP[76,77] | 0.018 | 0.028 | 0.018 | 0.031 |
| TPSS$_x$ | 0.017 | 0.036 | 0.017 | 0.039 |
| TPSS[78] | 0.014 | 0.027 | 0.014 | 0.030 |
| PBE0[14,15] | 0.023 | 0.025 | 0.022 | 0.025 |
| BHandHLYP[79] | 0.031 | 0.046 | 0.030 | 0.047 |
| B3LYP[80,81] | 0.016 | 0.022 | 0.015 | 0.023 |
| B2GPnoLYP | 0.044 | 0.063 | 0.043 | 0.064 |
| B2GP-PLYPnoPT2 | 0.042 | 0.062 | 0.041 | 0.063 |
| PW6B95[82] | 0.019 | 0.024 | 0.018 | 0.024 |
| M06[83] | 0.020 | 0.033 | 0.019 | 0.031 |
| M06-2X[83] | 0.026 | 0.039 | 0.023 | 0.038 |
| BMK[84] | 0.024 | 0.037 | 0.022 | 0.034 |
| LC-$\omega$HPBE[85] | 0.024 | 0.041 | 0.023 | 0.042 |
| CAM-B3LYP[86] | 0.022 | 0.032 | 0.020 | 0.032 |
| $\omega$B97X-D[87] | 0.020 | 0.029 | 0.019 | 0.028 |
| B2PLYP[88] | 0.011 | 0.015$^a$ | 0.010 | 0.016$^j$ |
| B2GP-PLYP[89] | 0.012 | 0.017$^b$ | 0.011 | 0.018$^k$ |
| DSD-PBEP86[90] | 0.010 | 0.017$^c$ | 0.009 | 0.019$^l$ |
| revDSD-PBEP86[61] | 0.009 | 0.017$^d$ | 0.008 | 0.019$^m$ |
| SOS0-PBE0-2[91] | 0.015 | 0.021$^e$ | 0.013 | 0.023$^n$ |
| PBE0-DH[92] | 0.021 | 0.024$^f$ | 0.020 | 0.024$^o$ |
| HF | 0.067 | 0.100$^g$ | 0.066 | 0.103$^p$ |
| MP2 | 0.011 | 0.065$^h$ | 0.011 | 0.074$^q$ |
| CCSD[93] | 0.012 | 0.021$^i$ | 0.012 | 0.022$^r$ |
| CCSD(T)[94,95] | REF | REF | REF | REF |

*Note:* The heat mapping for diagnostics within each column ranges from green (indicating the lowest RMSD) to red (representing the highest RMSD). The RMSD values after excluding $O_3$: $^a$0.014; $^b$0.015; $^c$0.012; $^d$0.010; $^e$0.016; $^f$0.024; $^g$0.095; $^h$0.031; $^i$0.020; $^j$0.015; $^k$0.015; $^l$0.013; $^m$0.011; $^n$0.016; $^o$0.023; $^p$0.096; $^q$0.034; $^r$0.021.

sets,[16] which are popular among many DFT practitioners because of their availability for nearly the entire Periodic Table.

## 3 | RESULTS AND DISCUSSION

In this manuscript, our aim is to investigate the performance of different density functional approximations (DFAs) as well as HF, MP2, CCSD and CCSD(T) methods for various types of partial charge analysis methods.

We will examine the distinctiveness of these methods, compare the consistency of various functionals within the Jacob's Ladder hierarchy, analyze the significance of semilocal and nonlocal correlation, and assess the influence of range separation. (For the avoidance of doubt: following the physics DFT literature (e.g., Reference 70 and other papers of the 'Perdew School') the term 'semilocal' refers here to dependence on density derivatives and 'nonlocal' to explicit dependence on occupied or virtual orbitals. In that sense, GGA and meta-GGA functionals are semilocal, while hybrids have fractions of nonlocal exchange, and double hybrids fractions of nonlocal correlation. This differs from some of the chemical literature where the term 'nonlocal' is used indiscriminately for all beyond-LDA functionals.) A key aspect of our study involves exploring the basis set convergence of various partial charge types. These findings will be beneficial for both developers and users, aiding in the selection of an appropriate theoretical level for accurate computation of partial charges.

The details of the partial charges obtained in this study are available in Microsoft Excel format as part of the Supporting Information.

## 3.1 | Selecting the optimal reference level of theory

Our first objective is to determine the most suitable reference level of theory. CCSD(T) is broadly considered to be the 'gold standard' of electronic structure methods (term first used by T. H. Dunning, Jr. in a 2000 lecture). Ideally, we would have used CCSD(T) as our reference level throughout; however, this proved computationally intractable for the larger basis sets. We are, however, able to calculate NPA charges at the CCSD(T)/haVQZ+d level for all closed-shell molecules of W4-17 except for the very largest species such as $C_2Cl_6$, $C_2F_6$, and n-pentane (Table 1), thus enabling us to evaluate the performance of lower-cost approaches.

The majority of double hybrid density functionals (DHDFs) outperform not only DFAs on lower rungs of Jacob's Ladder,[60] but also MP2 and CCSD. The lowest RMSD from CCSD(T) for NPA charges, 0.009 a.u., is attained by revDSD-PBEP86; its predecessor DSD-PBEP86 is the second-best method with a marginally larger RMSD = 0.010 a.u., while MP2 and CCSD yield RMSDs of 0.011 and 0.014 a.u., respectively. As is commonly seen in, for example, noncovalent interaction energies—see Reference 71 for a review—MP2 outperforms CCSD, as MP2 benefits from error compensation between the twin neglected effects of ΔCCSD−MP2 and Δ(T), which cannot happen in CCSD.

Furthermore, we succeeded in obtaining Generalized Atomic Polar Tensor (GAPT) partial charges at the CCSD(T)/haVTZ+d level for 149 out of 152 species (the missing species being again n-pentane and $C_2X_6$, X = F, Cl, $C_2Cl_6$) and at the CCSD(T)/haVQZ+d level for 126 species, comparing them with lower-level approaches such as various DFAs, MP2, and CCSD (Table 1). The CCSD(T)/haVTZ+d versus CCSD(T)/haVQZ+d RMS difference was less than 0.003 a.u., hence our discussion will focus on haVTZ+d for which we have nearly the entire dataset. Again, semi-empirical double hybrids come closest to CCSD(T), with RMS differences of 0.015 a.u. for B2PLYP, 0.016 a.u. for revDSD-PBEP86, and 0.017 a.u. for DSD-PBEP86, and B2GP-PLYP. Of the remarkably poor 0.063 a.u. for MP2, about half is due just to ozone, for which strong static correlation causes the MP2 GAPT charges to have the wrong sign (!). $S_4$ and $S_3$ are similarly, but more weakly, affected for the same reasons. The double hybrids are more resilient (as is generally true for double hybrids vs. MP2[72]): without ozone, RMS is lowest for revDSD-PBEP86 at just 0.010 a.u., compared to 0.014 for B2PLYP and 0.015 for B2GP-PLYP, while CCSD remains essentially unchanged at 0.020 a.u., and MP2 drops to 0.031 a.u. The picture remains the same with the CCSD(T)/haVQZ+d subset data. In all, revDSD-PBEP86 appears to be superior to both MP2 and CCSD for both NPA and GAPT; as Gaussian is capable of analytical second derivatives including intensities for double hybrids, we selected revDSD-PBEP86 as our 'secondary standard', for which we were able to use basis sets as large as haV5Z+d for most charge types. Henceforth, throughout the remainder of the manuscript, partial charges computed at the revDSD-PBEP86/haV5Z+d level of theory will be used as the reference, unless indicated otherwise.

Before drawing overarching conclusions, we shall survey individual charge types. Three themes recur across all of them: (1) the great importance of nonlocal correlation if a high fraction of HF exchange is present (this is a *fortiori* true of WFT methods with a full HF reference); (2) the performance of empirical double hybrids in general, and of revDSD-PBEP86 in particular, rivals or exceeds that of the CCSD wavefunction method; (3) for semilocal functionals and hybrids, the impact of semilocal correlation depends on functional and charge type.

## 3.2 | Analyzing DFA's robustness in calculating partial charges based on the electron density: Hirshfeld, Hirshfeld-I, and QTAIM

### 3.2.1 | Fuzzy density partitioning: Original ('vanilla') Hirshfeld

We successfully determined iterative Hirshfeld (see next subsection) partial charges for all 152 species using the haV5Z+d basis set (Table 2); regular Hirshfeld charges are obtained as a by-product at the first iteration. The reference method and other semi-empirical double hybrids yield exceedingly similar results: RMSDs obtained from DSD-PBEP86, B2GP-PLYP, and B2PLYP are only 0.001, 0.002, and 0.002 a.u., respectively.

Hybrid functionals (e.g., B3LYP, PW6B95, CAM-B3LYP, ωB97X-D, PBE0, BMK, M06, and M06-2X) performed similarly to MP2 and double hybrids. On the other hand, GGAs (e.g., PBE and BLYP) and meta-GGA (e.g., TPSS) exhibited larger RMSD values (0.016, 0.012, and 0.011 a.u., respectively).

To study the impact of semilocal correlation on Hirshfeld charge calculations, we conducted additional calculations excluding the semilocal correlation components of PBE, BLYP, and TPSS-based DFAs (PBEx, B88x, and TPSSx, respectively). These calculations revealed that semilocal correlation is not crucial for Hirshfeld charges.

However, removing PT2 correlation from B2GP-PLYP (B2GP-PLYPnoPT2) had a much larger impact, RMSD increasing tenfold from 0.002 to 0.020 a.u. This emphasizes the importance of nonlocal correlation in this context.

**TABLE 2** Summary of root mean square deviations (RMSDs) for partial charge calculations for closed-shell molecules of the W4-17 thermochemistry benchmark set.

| | Hirshfeld | Hirshfeld-I | HLY | NPA | GAPT | | | QTAIM |
| --- | --- | --- | --- | --- | --- | --- | --- | --- |
| | | | | | RMSD | MAD | MAD/RMSD ×(5/4) | |
| SVWN5 | 0.023 | 0.059 | 0.028 | 0.047 | 0.044 | 0.030 | 0.86 | 0.091 |
| PBE$_x$ | 0.017 | 0.033 | 0.038 | 0.015 | 0.037 | 0.025 | 0.86 | 0.062 |
| PBE | 0.016 | 0.040 | 0.018 | 0.029 | 0.036 | 0.021 | 0.73 | 0.059 |
| PBE-PW91LDA | 0.010 | 0.024 | 0.026 | 0.013 | 0.032 | 0.021 | 0.79 | 0.061 |
| B88$_x$ | 0.015 | 0.033 | 0.039 | 0.014 | 0.038 | 0.027 | 0.88 | 0.063 |
| BLYP | 0.012 | 0.028 | 0.022 | 0.017 | 0.033 | 0.019 | 0.73 | 0.063 |
| TPSS$_x$ | 0.016 | 0.041 | 0.043 | 0.021 | 0.041 | 0.031 | 0.94 | 0.068 |
| TPSS | 0.011 | 0.027 | 0.016 | 0.014 | 0.032 | 0.020 | 0.77 | 0.046 |
| PBE0 | 0.006 | 0.017 | 0.009 | 0.016 | 0.028 | 0.014 | 0.62 | 0.021 |
| BHandHLYP | 0.013 | 0.027 | 0.021 | 0.026 | 0.048 | 0.026 | 0.67 | 0.026 |
| B3LYP | 0.004 | 0.010 | 0.014 | 0.009 | 0.027 | 0.012 | 0.56 | 0.032 |
| B2GPnoLYP | 0.020 | 0.048 | 0.035 | 0.041 | 0.065[a] | 0.039[a] | 0.75 | 0.058 |
| B2GP-PLYPnoPT2 | 0.020 | 0.043 | 0.029 | 0.038 | 0.063[a] | 0.036[a] | 0.71 | 0.049 |
| PW6B95 | 0.004 | 0.010 | 0.008 | 0.012 | 0.028 | 0.012 | 0.54 | 0.023 |
| M06 | 0.006 | 0.010 | 0.011 | 0.013 | 0.033 | 0.017 | 0.64 | 0.043 |
| M06-2X | 0.006 | 0.013 | 0.012 | 0.018 | 0.041 | 0.021 | 0.63 | 0.019 |
| BMK | 0.006 | 0.013 | 0.011 | 0.016 | 0.037 | 0.021 | 0.70 | 0.023 |
| LC-$\omega$PBE | 0.009 | 0.022 | 0.014 | 0.018 | 0.043 | 0.025 | 0.72 | 0.028 |
| CAM-B3LYP | 0.005 | 0.012 | 0.013 | 0.014 | 0.034 | 0.017 | 0.62 | 0.027 |
| $\omega$B97X-D | 0.005 | 0.012 | 0.008 | 0.013 | 0.031 | 0.014 | 0.56 | 0.021 |
| B2PLYP | 0.002 | 0.003 | 0.007 | 0.003 | 0.012[a] | 0.005[a] | 0.53 | 0.015 |
| B2GP-PLYP | 0.002 | 0.005 | 0.005 | 0.004 | 0.009[a] | 0.006[a] | 0.86 | 0.004 |
| DSD-PBEP86 | 0.001 | 0.002 | 0.002 | 0.002 | 0.003[a] | 0.002[a] | 0.90 | 0.003 |
| revDSD-PBEP86 | REF | REF | REF | REF | REF | REF | | REF |
| SOS0-PBE0-2 | 0.005 | 0.010 | 0.006 | 0.007 | 0.011[a] | 0.007[a] | 0.78 | 0.016 |
| PBE0-DH | 0.005 | 0.010 | 0.011 | 0.013 | 0.026[a] | 0.014[a] | 0.69 | 0.012 |
| HF | 0.034 | 0.076 | 0.045 | 0.064 | 0.102 | 0.059 | 0.72 | 0.105 |
| MP2 | 0.004 | 0.010 | 0.008 | 0.009 | 0.050 | 0.016 | 0.39 | 0.017 |
| CCSD | 0.007[b] | 0.016[b] | 0.012[b] | 0.011[b] | 0.029[b] | | | 0.020[b] |
| Variable scales for perspective (revDSD-PBEP86/haV5Z+d) | | | | | | | | |
| $\|q\|_1$ (mean abs.) | 0.100 | 0.272 | 0.228 | 0.344 | | 0.310 | | 0.494 |
| $\|q\|_2$ (rms) | 0.139 | 0.399 | 0.300 | 0.462 | 0.490 | | | 0.781 |
| PC$_1$ (tab. 4, Reference 12) | 0.085 | 0.278 | 0.229 | 0.304 | 0.258 | | | 0.419 |

*Note:* The heat mapping for diagnostics within each column ranges from green (indicating the lowest RMSD) to red (representing the highest RMSD). The haV5Z+d basis set is used, unless specified otherwise.

[a]The statistics correspond to the haVQZ+d basis set (reference: revDSD-PBEP86/haVQZ+d).

[b]The statistics correspond to the haVTZ+d basis set (reference: revDSD-PBEP86/haVTZ+d).

### 3.2.2 | Fuzzy density partitioning redux: Iterative Hirshfeld

Moving on to Hirshfeld-I, the RMSDs are three times larger than what we observed for ordinary Hirshfeld, which is not surprising as this is also true of the charges themselves. As expected, DSD-PBEP86, B2PLYP, and B2GP-PLYP DFAs again stayed closest to the reference level, deviating by 0.002, 0.003, and 0.005 a.u. RMS, respectively (Table 2).

Descending a rung on Jacob's Ladder, the PW6B95, B3LYP, and M06 global hybrids are tied there for lowest RMSD (0.010 a.u.).

10969987a, 2024, 11. Downloaded from https://onlinelibrary.wiley.com/doi/10.1002/jcc.27294 by University Of Florida, Wiley Online Library on [04/04/2024]. See the Terms and Conditions (https://onlinelibrary.wiley.com/terms-and-conditions) on Wiley Online Library for rules of use; OA articles are governed by the applicable Creative Commons License

Surprisingly, range-separated hybrids seemingly under-perform their global-hybrid counterparts; for example, CAM-B3LYP and LC-$\omega$hPBE have RMSDs of 0.012 and 0.022 a.u., compared to 0.010 and 0.017 a. u., respectively, for the corresponding global hybrids B3LYP and PBE0.

What happens when semilocal correlation is removed depends on the functional. PBEx actually outperforms PBE (0.033 vs. 0.040 a.u.), but TPSSx is inferior to TPSS (0.041 vs. 0.027 a.u.), while B88x and BLYP are in a statistical dead heat (0.033 vs. 0.028 a.u.).

When we removed PT2 nonlocal correlation from the B2GP-PLYP functional, we again observed degradation by about an order of magnitude, with RMSD increasing from 0.005 to 0.043 a.u. In addition removing the LYP correlation hardly matters in comparison (RMSD of 0.048 a.u. for the B2GPnoLYP variant). It would seem that when a high fraction of HF exchange is present, nonlocal correlation is essential; this is further confirmed by comparing HF (0.076 a.u.) with MP2 (0.010 a.u.). Incidentally, SVWN5 (i.e., LDA) performs almost as poorly as HF, with an RMSD of 0.059 a.u.

### 3.2.3 | Disjoint charge partitioning: QTAIM ("Bader charges")

Turning to QTAIM charges, we observed that the RMSDs are nearly quadruple those of the previously discussed methods. (This time, however, it cannot be all attributed to QTAIM charges *themselves* just being larger.)

Some species (specifically, $C_2Cl_2$, $C_2H_2$, $C_2ClH$, FCCF, $Si_2H_6$, and propyne) exhibited non-nuclear attractors at one or more levels of theory (NNAs, i.e., pseudo-atomic density basins that do not contain a nucleus). We ruled out that any of these are artifacts of the grid used for locating the zero-flux surfaces by rerunning AIMALL with extremely fine grids. The origin of NNAs has been discussed in several studies,[96–98] most recently Reference [99]. Intriguingly, many (but not all) NNAs disappear as the basis set is expanded; in addition, we observed that pure DFAs (i.e., GGAs and mGGAs) appear to be less prone to NNAs than WFT methods.

In order to put statistics for QTAIM on an equal footing, we have eliminated any species that exhibited NNAs at any level. As allene (with its cumulene double bonds) exhibited the same erratic behavior, we eliminated that as well. HF has the largest RMSD of all approaches, 0.105 a.u.; with an RMSD=0.091, SVWN5 (i.e., LDA, rung one) comes in second place. MP2 cuts down RMSD to 0.017 a.u.

GGAs (i.e., rung 2 on the Jacob's ladder), like PBE and BLYP, rendered RMSDs of 0.059 and 0.063, respectively. The omission of semilocal correlation from these functionals had an insignificant impact.

TPSS yields RMSD = 0.046 a.u., which increases to 0.068 a.u. when the semilocal correlation is suppressed—and as we just saw the same for the Hirshfeld variants, this may indicate a nontrivial significance of semilocal correlation for meta-GGAs.

M06-2X exhibits the smallest deviation among hybrid DFAs (RMSD = 0.019 a.u.), followed by PBE0 and $\omega$B97X-D, both having

RMSD = 0.021 a.u. PW6B95 and BMK take third and fourth place, each exhibiting RMSDs of 0.023 a.u.

Interestingly, B3LYP, a hybrid DFA that did very well for regular and iterative Hirshfeld performs much less well (0.032 a.u.) for QTAIM.

We note in passing that switching from B3LYP to CAM-B3LYP reduces the deviation from 0.032 to 0.027 a.u., while shifting from PBE0 to LC-$\omega$hPBE increases the deviation from 0.021 to 0.028 a.u.

DSD-PBEP86 and B2GP-PLYP unsurprisingly come quite close to revDSD-PBEP86, with RMSDs of only 0.003 and 0.004 a.u., respectively. B2PLYP however displayed a rather larger RMSD of 0.015 a.u. To our surprise, with RMSD = 0.012 a.u., PBE0-DH—a non-empirical double hybrid that performed fairly poorly for the Hirshfeld variants—outperformed SOS0-PBE0-2 and B2PLYP double hybrids.

Furthermore, removing the PT2 correlation from the B2GP-PLYP functional once again increased RMSD by an order of magnitude, re-emphasizing the critical importance of nonlocal correlation. Once again, additionally eliminating the semilocal correlation component had a negligible further impact (the RMSD rose by just 0.009 a.u.).

To our surprise, M06 yields unexpectedly high errors (RMSD=0.043 a.u.) compared to its sibling M06-2X (RMSD = 0.019 a. u.). These findings persisted even when we changed the basis set to haVQZ+d or haVTZ+d, or haVDZ+d or increased the grid size to superfine.

## 3.3 | Atomic orbital-based: Natural population analysis

When examining the NPA charges obtained from NBO7 (Natural Bond Orbital analysis), we find striking similarities to the observations obtained for iterative Hirshfeld (Table 2). For instance, the RMSD decreases from 0.029 to 0.015 a.u. when the semilocal correlation from PBE is turned off, while the RMSD for TPSS DFA increases from 0.014 to 0.021 a.u. On the one hand, these similarities might seem surprising given the very different physical foundations of the charge types; on the other hand, CSESEM[12] found $R^2 = 0.92$ between NPA and Hirshfeld-I for nearly the whole GMTKN55 set.

All semi-empirical double hybrids, as well as the nonempirical hybrid SOS0-PBE0-2, exhibit smaller errors (0.002–0.004 a.u.) than MP2 (0.009 a.u.).

In what is becoming a pattern across all charge types, switching off nonlocal PT2 correlation in B2GP-PLYP degrades RMSD by an order of magnitude.

In addition, both HF and SVWN5 exhibit poor performance, with RMSDs of 0.064 and 0.047 a.u., respectively.

## 3.4 | Electrostatic potential charges: Hu-Lu-Yang as an example

The obvious winners here are semi-empirical double hybrids along with SOS0-PBE0-2 (a non-empirical DHDF). For example, the RMSDs for DSD-PBEP86, B2GP-PLYP, SOS0-PBE0-2, and B2PLYP are just 0.002, 0.005, 0.006, and 0.007 a.u., respectively, which are all lower than the RMSD value of 0.008 a.u. for MP2 (Table 2).

PW6B95 and $\omega$B97X-D are tied for the lowest deviation (RMSD = 0.008 a.u.) among the hybrid DFAs, with PBE0 coming in second at 0.009 a.u.

Lower-rung DFAs, such as PBE, BLYP, and TPSS, have RMSDs of 0.018, 0.022, and 0.016 a.u., respectively.

Unlike vanilla and iterative Hirshfeld charges, HLY errors increase by a factor of two when the semilocal correlation is removed from the PBE functional. Likewise, deviations increase from 0.022 to 0.039 a.u. for the BLYP functional and from 0.016 to 0.043 a.u. for the TPSS functional, indicating that semilocal correlation is necessary for ESP-based charges.

Continuing the trend from the other charges in an attenuated form, the statistical accuracy decreases by a factor of 'only' five when the nonlocal correlation is omitted from the B2GP-PLYP functional, and again this dwarfs the impact of semilocal correlation (RMSDs are 0.029 and 0.035 a.u., respectively, for B2GP-PLYP-noPT2 and B2GPnoLYP).

With RMSDs of 0.028 and 0.045 a.u., respectively, SVWN5 performs slightly better than HF.

Electrostatic potential charge fits on organic and biological macromolecules run into a variety of numerical problems, in response to which the RESP (restrained electrostatic potential) approach[100] was developed by the Kollman[101] (AMBER) group. In its simplest form, a hyperbolic penalty function of the form $a\sum_j(\sqrt{q_j^2+b^2}-b)$ is added to the least-squares fitting function, where the $q_i$ are the partial charges sought and a,b are parameters. Additional constraints can be added to ensure that equivalent atoms have the same charges. A reviewer requested that we consider RESP. While (on account of the parameters) RESP charges are not uniquely defined, we carried out exploratory calculations at the HF/haVTZ+d and PBE0/haVTZ+d levels using the RESP implementation in Multiwfn[58] with default parameters, comparing them to Merz-Kollman[24] charges (to which RESP reduces for $a=0$.) While this revealed significant differences between RESP and ESP (MK) for propane, n-butane, n-pentane, and similar molecules, the basis set convergence behavior was essentially the same as for the HLY reported in the present paper.

## 3.5 | GAPT (atomic polar tensor) charges

Let us turn our focus to the GAPT partial charges (Table 2). Due to the high computational resource requirements (effectively those for a frequency calculation), we were unable to calculate GAPT charges using the hAV5Z+d basis set for all DFAs. We performed GAPT charge computations for all 152 species using the haV5Z+d basis set for up to the fourth rung of Jacob's Ladder, of which HF (also included) can be seen as a special case. Among fifth-rung DFAs, GAPT charge calculations were performed only for the revDSD-PBEP86 functional on the complete set. At the MP2/haV5Z+d level, we successfully obtained GAPT charges for all species except $C_2Cl_6$.

HF yields the poorest performance with an RMSD of 0.102 a.u., followed by MP2 in second place with an RMSD of 0.050 a.u. Closer inspection reveals that the MP2 statistic is being skewed by large errors for systems exhibiting strong static correlation—the partial charges in ozone even reverse sign! (Repeating these calculations at the CASSCF level with various active spaces from (2/2) to full valence, we obtain charges qualitatively similar to CCSD and CCSD(T) but not to MP2.)

Given that mean absolute deviations (MADs) are more robust to outliers, it is worth considering MADs herein. MAD/RMSD ratios much below the ideal value for an unbiased normal distribution[55,102] of $\sqrt{2/\pi}\approx4/5$ indicate the presence of outliers. For instance, the MAD/RMSD ratio for MP2 is 0.32, which reflects that the unusually large RMSD yielded by MP2 for GAPT charges is due to a handful of species rather than MP2 being broadly inadequate for GAPT charges.

SVWN5 produces an RMSD of 0.044 a.u., while BLYP and PBE yield slightly lower values of 0.033 and 0.036 a.u., respectively. TPSS shows the RMSD at 0.032 a.u.

Removing semilocal correlation from GGAs has minimal impact on the results. For PBE and BLYP, RMSDs shift from 0.036 to 0.037 a.u. and 0.033 to 0.038 a.u., respectively. However, TPSS exhibits a more profound increase in RMSD, rising from 0.032 to 0.041 a.u.

Among hybrid DFAs, B3LYP exhibits the smallest RMSD, closely followed by PBE0 and PW6B95. The inclusion of range-separation once again appears to have an adverse effect, resulting in an increase in RMSD from 0.028 to 0.043 a.u. when shifting from PBE0 to LC-$\omega$hPBE.

For the haVQZ+d basis set, we have the complete set of results for all DHDFs. The RMS change between haVQZ+d and haV5Z+d basis sets for revDSD-PBEP86 is just a negligible 0.0004 a.u. (Table 3), which corroborates that our results at the haVQZ+d level are adequately converged with the basis set.

Continuing the trend, semi-empirical double hybrids and SOS0-PBE0-2 are the winners, led by DSD-PBEP86 with an RMSD of only 0.003 a.u., B2GP-PLYP with an RMSD of 0.009 a.u.is at the second position, and B2PLYP and SOS0-PBE0-2 are tied for third place with RMSD = 0.012 a.u.

And continuing another trend, the vital importance of nonlocal correlation is evident by a substantial increase in RMSD (from 0.010 to 0.063) when the PT2 correlation is removed in B2GP-PLYP.

## 3.6 | Basis set convergence

Having discussed the behavior of different functionals for different charges near the basis set limit, we shall now turn to their basis set convergence. In this section, we shall see that Hirshfeld, Hirshfeld-I, HLY, and GAPT methods exhibit rapid basis set convergence, in contrast to NPA and QTAIM.

The basis set convergence of partial charge calculations is presented in Table 3 as a sequence of RMS basis set increments, starting from haVDZ+d→haVTZ+d, then proceeding via haVTZ+d→haVQZ

+d to haVQZ+d→haV5Z+d. Residual basis set increments below RMSD = 0.001 a.u. can be considered negligible.

Let us start by discussing ordinary Hirshfeld charges. The RMSDs are small, ranging from 0.002 to 0.005 a.u. when turning from haVDZ +d→haVTZ+d, from 0.000 to 0.002 a.u. when proceeding from haVTZ+d→haVQZ+d, and from 0.000 to 0.001 a.u. when switching from haVQZ+d to haV5Z+d. This indicates that the haVTZ+d basis set or just even haVDZ+d is sufficient for conducting ordinary Hirshfeld charge calculations.

It should be noted that, for technical reasons, our partial charge calculations at the CCSD level are limited to the haVDZ+d and haVTZ +d basis sets, with the exception of NBO7, which we were able to compute using MOLPRO.

Turning to Hirshfeld-I charges, we observed that the RMSDs vary from 0.005 to 0.014 a.u. when moving from haVDZ+d to haVTZ+d, nearly three times larger than what we observed for ordinary Hirshfeld, but similar in relative terms (as Hirshfeld-I charges tend to be $\simeq 3\times$ as big). For the HF and MP2 methods, we obtained RMSD values of 0.006 and 0.013 a.u., respectively. The correlation component of MP2 (i.e., HF-MP2) contributes an RMSD of 0.010 a.u., signifying that MP2 correlation exhibits slower basis set convergence than Hartree-Fock. Additionally, the CCSD method yields an RMSD of 0.014 a.u., where CCSD-HF results in an RMSD of 0.011 a.u. This underscores that CCSD correlation also exhibits slower basis set convergence when compared to HF. The RMSDs for MP2's correlation dwindled to 0.004 and 0.001 a.u. when progressing from haVTZ+d to haVQZ+d and from haVQZ+d to haV5Z+d, respectively. Furthermore, it is worth noting that the majority of the examined methods yielded RMSDs within the range of 0.001 a.u. for both the haVTZ+d to haVQZ+d and haVQZ+d to haV5Z+d basis set increments. This suggests that the haVTZ+d basis set can be deemed adequate for the determination of Hirshfeld-I charges.

Turning our attention to the HLY electrostatic charges, we observe that HF results in a relatively substantial RMSD of 0.018 a.u. for haVDZ+d→haVTZ+d. MP2 exhibits an RMSD of 0.015 a.u. (likely due to some error cancellation), while CCSD results in an RMSD of 0.019 a.u. A detailed evaluation reveals that basis set increments for both MP2 and CCSD's correlation components hover around 0.01 a. u., falling within a similar range as HF. This suggests that the slow basis set convergence observed at the haVDZ+d→haVTZ+d level stems from both the HF and correlation components. Furthermore, for the haVTZ+d→haVQZ+d, most methods consistently show RMSDs within the 0.002–0.003 a.u. range. Similarly, when employing haVQZ+d→haV5Z+d, the majority of methods yield negligible RMSDs around 0.001 a.u. This shows that haVTZ+d or *a fortiori* haVQZ+d are suitable for HLY charge calculations.

In congruence with the ordinary Hirshfeld, Hirshfeld-I, and HLY charge schemes, GAPT also exhibits rapid basis set convergence. For instance, HF yields an RMSD of 0.005 a.u. at the haVDZ+d→haVTZ +d level, which is found to decrease to 0.0008 and 0.0002 a.u. for haVTZ+d→haVQZ+d and haVQZ+d→haV5Z+d, respectively. Furthermore, MP2 and CCSD's correlation contributions were found to have similar basis set dependence as HF, as can be seen in Table 3.

Turning now to NPA: its basis set convergence seems to be slower than ordinarily Hirshfeld, Hirshfeld-I, HLY, and GAPT methods, yet faster than QTAIM. However, we should keep in mind that QTAIM charges are larger than NPA on average—the RMS charges at the revDSD-PBEP86/haV5Z+d level are {0.462, 0.781} for {NPA, QTAIM}, while as an alternative scale gauge, the 'loadings' (i.e., coefficients) in the first principal component of ionicity in tab. 4 of Reference 12 are {0.304, 0.409}. Taking this into account, NPA and QTAIM actually exhibit fairly similar basis set convergence in a *relative* sense.

Expanding the basis set from haVDZ+d to haVTZ+d leads to RMSDs ranging from 0.033 to 0.041 a.u., with most density functional approximations showing deviations in the vicinity of 0.03–0.04 a.u. Notably, HF exhibits an RMSD of 0.038 a.u., while the correlation components of MP2 and CCSD exhibit RMSDs of 0.006 and 0.007 a. u., respectively. This underscores that the slow basis set convergence observed in NPA charges can be entirely attributed to the HF component. Similarly, LDA also demonstrates a slow basis set convergence akin to HF.

As mentioned earlier, we successfully computed NPA charges at the CCSD(T) level for a subset of molecules using haVDZ+d, haVTZ +d and haVQZ+d basis sets. It is noteworthy that the triples component (i.e., CCSD(T)-CCSD)) demonstrates rapid basis set convergence, with RMSDs of only 0.002 a.u. each for haVDZ+d to haVTZ+d and haVTZ+d to haVQZ+d.

Hartree-Fock's RMS basis set increment is halved upon increasing the basis set size to {haVTZ+d, haVQZ+d}, reaching 0.014 a.u., while the corresponding value for MP2's component decreases to 0.002 a.u. Surprisingly, even with a basis set pair as extensive as {haVQZ+d, haV5Z+d}, HF's RMSD remains at 0.012 a.u. This pattern is consistently observed across all tested methods. A closer examination of individual systems reveals that these differences are not limited to just a few cases.

QTAIM exhibits the strongest basis set dependence in absolute terms: see however the remark above about relative magnitudes. The RMSD basis set increments from haVDZ+d to haVTZ+d fall in the 0.048–0.065 a.u. range, for the haVTZ+d/haVQZ+d pair in the 0.019–0.044 range. When proceeding from haVQZ+d to the largest available basis set, haV5Z+d, the RMSD range from 0.012 to 0.019 a.u. Furthermore, HF exhibits an RMS increment of 0.065 a.u., whereas the MP2 and CCSD correlation components display RMSDs of 0.024 and 0.029 a.u., respectively, showing that HF is responsible for the slow overall basis set convergence. A detailed examination of individual molecules revealed that the disparities are pervasive rather than confined to just a select few systems. Nevertheless, it is noteworthy that $C_2ClH_3$, $C_2H_3F$, $CH_2$, and $H_2S$ exhibited the highest deviations for HF. Although marginally lower than HF, we should point out that all DFAs also exhibited RMSDs in the vicinity of 0.05 a.u.

Expanding the basis set size to {haVTZ+d, hAVQZ+d} resulted in a reduction of errors to approximately 0.03 a.u. for DFAs, which, while improved, remain annoying. For HF, the RMSD stands at 0.044 a.u. Interestingly, the RMSD for MP2-HF at the {hAVTZ+d, hAVQZ

**TABLE 3**   Basis set convergence (RMSDs) for partial charge calculations for closed-shell molecules of the W4-17 thermochemistry benchmark set.

| | haVDZ+d to haVTZ+d | | | | | |
| --- | --- | --- | --- | --- | --- | --- |
| | Hirshfeld | Hirshfeld-I | HLYgat | NBO7 | GAPT | QTAIM |
| SVWN5 | 0.002 | 0.006 | 0.017 | 0.036 | 0.005 | 0.050 |
| PBE$_x$ | 0.002 | 0.006 | 0.021 | 0.034 | 0.006 | 0.049 |
| PBE | 0.002 | 0.006 | 0.018 | 0.035 | 0.004 | 0.049 |
| B88$_x$ | 0.002 | 0.005 | 0.019 | 0.035 | 0.005 | 0.049 |
| BLYP | 0.002 | 0.006 | 0.020 | 0.035 | 0.005 | 0.050 |
| TPSS$_x$ | 0.002 | 0.006 | 0.019 | 0.034 | 0.006 | 0.048 |
| TPSS | 0.002 | 0.006 | 0.017 | 0.036 | 0.005 | 0.048 |
| PBE0 | 0.002 | 0.006 | 0.017 | 0.037 | 0.004 | 0.050 |
| BHandHLYP | 0.002 | 0.005 | 0.017 | 0.037 | 0.005 | 0.051 |
| B3LYP | 0.002 | 0.005 | 0.018 | 0.036 | 0.005 | 0.049 |
| B2GPnoLYP | 0.002 | 0.005 | 0.017 | 0.037 | 0.004 | 0.055 |
| B2GP-PLYPnoPT2 | 0.002 | 0.005 | 0.017 | 0.037 | 0.005 | 0.054 |
| PW6B95 | 0.002 | 0.006 | 0.018 | 0.035 | 0.004 | 0.049 |
| M06 | 0.003 | 0.008 | 0.015 | 0.041 | 0.008 | 0.049 |
| M06-2X | 0.003 | 0.007 | 0.017 | 0.036 | 0.007 | 0.048 |
| BMK | 0.003 | 0.006 | 0.013 | 0.039 | 0.010 | 0.051 |
| LC-$\omega$hPBE | 0.002 | 0.006 | 0.018 | 0.035 | 0.004 | 0.052 |
| CAM-B3LYP | 0.002 | 0.006 | 0.018 | 0.036 | 0.005 | 0.050 |
| $\omega$B97X-D | 0.002 | 0.006 | 0.015 | 0.037 | 0.004 | 0.049 |
| B2PLYP | 0.002 | 0.007 | 0.016 | 0.035 | 0.004 | 0.050 |
| B2GP-PLYP | 0.003 | 0.007 | 0.016 | 0.035 | 0.004 | 0.052 |
| DSD-PBEP86 | 0.003 | 0.008 | 0.015 | 0.036 | 0.004 | 0.053 |
| revDSD-PBEP86 | 0.003 | 0.008 | 0.015 | 0.036 | 0.004 | 0.053 |
| SOS0-PBE0-2 | 0.003 | 0.009 | 0.015 | 0.036 | 0.004 | 0.056 |
| PBE0-DH | 0.002 | 0.006 | 0.017 | 0.037 | 0.003 | 0.052 |
| HF | 0.002 | 0.006 | 0.018 | 0.038 | 0.005 | 0.065 |
| MP2 | 0.004 | 0.013 | 0.015 | 0.034 | 0.006 | 0.057 |
| MP2-HF | 0.003 | 0.010 | 0.013 | 0.006 | 0.008 | 0.029 |
| CCSD | 0.005 | 0.014 | 0.019 | 0.033 | 0.005 | 0.060 |
| CCSD-HF | 0.004 | 0.011 | 0.014 | 0.007 | 0.006 | 0.024 |
| CCSD(T) | | | | 0.033 | 0.006 | |
| CCSD(T)-CCSD | | | | 0.002 | 0.002 | |

| | haVTZ+d to haVQZ+d | | | | | |
| --- | --- | --- | --- | --- | --- | --- |
| | Hirshfeld | Hirshfeld-I | HLYgat | NBO7 | GAPT | QTAIM |
| SVWN5 | 0.001 | 0.001 | 0.003 | 0.016 | 0.001 | 0.030 |
| PBE$_x$ | 0.001 | 0.001 | 0.002 | 0.015 | 0.001 | 0.029 |
| PBE | 0.001 | 0.001 | 0.002 | 0.016 | 0.001 | 0.030 |
| B88$_x$ | 0.001 | 0.001 | 0.003 | 0.015 | 0.001 | 0.028 |
| BLYP | 0.001 | 0.001 | 0.002 | 0.015 | 0.001 | 0.029 |
| TPSS$_x$ | 0.001 | 0.001 | 0.002 | 0.015 | 0.001 | 0.033 |
| TPSS | 0.001 | 0.001 | 0.003 | 0.016 | 0.001 | 0.034 |
| PBE0 | 0.001 | 0.001 | 0.002 | 0.015 | 0.001 | 0.031 |

**TABLE 3** (Continued)

| | haVTZ+d to haVQZ+d | | | | | |
|---|---|---|---|---|---|---|
| | Hirshfeld | Hirshfeld-I | HLYgat | NBO7 | GAPT | QTAIM |
| BHandHLYP | 0.001 | 0.001 | 0.003 | 0.015 | 0.001 | 0.032 |
| B3LYP | 0.001 | 0.001 | 0.003 | 0.015 | 0.001 | 0.030 |
| B2GPnoLYP | 0.001 | 0.001 | 0.003 | 0.015 | 0.001 | 0.034 |
| B2GP-PLYPnoPT2 | 0.001 | 0.001 | 0.003 | 0.015 | 0.001 | 0.034 |
| PW6B95 | 0.001 | 0.001 | 0.002 | 0.015 | 0.001 | 0.028 |
| M06 | 0.002 | 0.004 | 0.007 | 0.017 | 0.007 | 0.019 |
| M06-2X | 0.000 | 0.001 | 0.005 | 0.016 | 0.003 | 0.035 |
| BMK | 0.001 | 0.003 | 0.007 | 0.017 | 0.005 | 0.033 |
| LC-$\omega$hPBE | 0.001 | 0.001 | 0.003 | 0.015 | 0.001 | 0.032 |
| CAM-B3LYP | 0.001 | 0.001 | 0.003 | 0.015 | 0.001 | 0.031 |
| $\omega$B97X-D | 0.001 | 0.001 | 0.003 | 0.016 | 0.001 | 0.029 |
| B2PLYP | 0.001 | 0.002 | 0.002 | 0.015 | 0.001 | 0.030 |
| B2GP-PLYP | 0.001 | 0.002 | 0.002 | 0.015 | 0.001 | 0.030 |
| DSD-PBEP86 | 0.001 | 0.002 | 0.003 | 0.015 | 0.001 | 0.030 |
| revDSD-PBEP86 | 0.001 | 0.002 | 0.003 | 0.015 | 0.001 | 0.030 |
| SOS0-PBE0-2 | 0.001 | 0.002 | 0.006 | 0.017 | 0.002 | 0.031 |
| PBE0-DH | 0.001 | 0.001 | 0.003 | 0.015 | 0.001 | 0.032 |
| HF | 0.001 | 0.001 | 0.003 | 0.014 | 0.001 | 0.044 |
| MP2 | 0.002 | 0.004 | 0.003 | 0.015 | 0.002 | 0.029 |
| MP2-HF | 0.000 | 0.004 | 0.002 | 0.002 | 0.002 | 0.022 |
| CCSD | | | | 0.015 | 0.003 | |
| CCSD-HF | | | | 0.003 | 0.003 | |
| CCSD(T) | | | | 0.014 | 0.003 | |
| CCSD(T)-CCSD | | | | 0.002 | 0.000 | |

| | haVQZ+d to haV5Z+d | | | | | |
|---|---|---|---|---|---|---|
| | Hirshfeld | Hirshfeld-I | HLYgat | NBO7 | GAPT | QTAIM |
| SVWN5 | 0.000 | 0.001 | 0.001 | 0.013 | 0.000 | 0.012 |
| PBE$_x$ | 0.000 | 0.001 | 0.002 | 0.013 | 0.001 | 0.016 |
| PBE | 0.000 | 0.001 | 0.001 | 0.013 | 0.000 | 0.015 |
| B88$_x$ | 0.000 | 0.001 | 0.001 | 0.013 | 0.001 | 0.015 |
| BLYP | 0.000 | 0.001 | 0.001 | 0.013 | 0.000 | 0.015 |
| TPSS$_x$ | 0.000 | 0.001 | 0.001 | 0.013 | 0.001 | 0.013 |
| TPSS | 0.000 | 0.001 | 0.001 | 0.013 | 0.000 | 0.013 |
| PBE0 | 0.000 | 0.000 | 0.001 | 0.012 | 0.000 | 0.015 |
| BHandHLYP | 0.000 | 0.001 | 0.001 | 0.012 | 0.000 | 0.014 |
| B3LYP | 0.000 | 0.001 | 0.001 | 0.013 | 0.000 | 0.015 |
| B2GPnoLYP | 0.000 | 0.000 | 0.001 | 0.012 | | 0.014 |
| B2GP-PLYPnoPT2 | 0.000 | 0.000 | 0.001 | 0.012 | | 0.014 |
| PW6B95 | 0.000 | 0.001 | 0.001 | 0.013 | 0.000 | 0.014 |
| M06 | 0.001 | 0.002 | 0.003 | 0.013 | 0.007 | 0.017 |
| M06-2X | 0.001 | 0.001 | 0.002 | 0.013 | 0.002 | 0.019 |
| BMK | 0.001 | 0.001 | 0.003 | 0.013 | 0.002 | 0.013 |
| LC-$\omega$hPBE | 0.000 | 0.001 | 0.001 | 0.013 | 0.000 | 0.016 |
| CAM-B3LYP | 0.000 | 0.001 | 0.001 | 0.013 | 0.000 | 0.015 |

**TABLE 3** (Continued)

| | haVQZ+d to haV5Z+d | | | | | |
|---|---|---|---|---|---|---|
| | Hirshfeld | Hirshfeld-I | HLYgat | NBO7 | GAPT | QTAIM |
| ωB97X-D | 0.000 | 0.001 | 0.001 | 0.013 | 0.000 | 0.016 |
| B2PLYP | 0.000 | 0.001 | 0.001 | 0.013 | | 0.013 |
| B2GP-PLYP | 0.000 | 0.001 | 0.001 | 0.013 | | 0.013 |
| DSD-PBEP86 | 0.000 | 0.001 | 0.001 | 0.012 | | 0.013 |
| revDSD-PBEP86 | 0.000 | 0.001 | 0.001 | 0.012 | 0.000 | 0.013 |
| SOS0-PBE0-2 | 0.000 | 0.001 | 0.001 | 0.012 | | 0.013 |
| PBE0-DH | 0.000 | 0.000 | 0.001 | 0.012 | | 0.014 |
| HF | 0.000 | 0.000 | 0.001 | 0.012 | 0.000 | 0.014 |
| MP2 | 0.001 | 0.001 | 0.001 | 0.013 | | 0.012 |
| MP2-HF | 0.001 | 0.001 | 0.001 | 0.002 | 0.001 | 0.005 |

*Note:* The heat mapping for diagnostics within each column ranges from green (indicating the lowest RMSD) to red (representing the highest RMSD). Heat mapping is done separately within each column for each basis set pair.

+d} level falls within the same range as that for {hAVDZ+d, hAVTZ +d}, which is somewhat unexpected. Upon conducting an in-depth analysis of individual systems, it becomes evident that $C_2ClH_3$, $C_2H_3F$, $CH_2C$, and $H_2S$ predominantly account for these discrepancies. Note that the latter persist even with larger grid settings and when using the different QTAIM module in MULTIWFN, thus ruling out a code or grid artifact. Expanding the basis set to {haVQZ+d, haV5Z+d} yields an RMSD of approximately 0.015 a.u., representing a reduction in error. For HF, the RMSD stands at 0.014 a.u. Notably, the MP2 correlation contribution exhibits a reduced RMSD of 0.005 a.u., which aligns the expected trend of error reduction with increasing basis set size. This suggests that anomalies are likely encountered with smaller basis sets, such as haVTZ+d, for certain species, and thus underscores the critical importance of utilizing larger basis sets whenever feasible in QTAIM calculations. (As this paper was being finalized for submission, a study by Santos et al.[103] was published, which showed the much reduced basis set sensitivity of overlap properties descriptors compared to QTAIM.)

Finally, we can state overall that our findings on the performance of DFT and WFT methods remain consistent, regardless of basis set size.

## 3.7 | A note on the performance of 'calendar' basis sets

At the request of a reviewer, we explored the performance of so-called 'calendar' basis sets,[67,68] where 'aug-cc-pV(n+d)Z' corresponds to the fully augmented basis set (including hydrogen), 'jul-cc-pV(n+d)Z' omits diffuse functions on hydrogen (and is hence equivalent to aug'-cc-pV(n+d)Z in Del Bene's notation[104] or heavy-aug-cc-pV(n+d)Z or haVnZ+d in Hobza's notation,[71] that is, what we have been using thus far), and jun-cc-pV(n+d)Z also omits the diffuse functions with the highest angular momenta on non-hydrogen atoms. (may-cc-pV(n +d)Z omits the diffuse functions with the two *two* angular momenta, and so forth). We only explored this for revDSD-PBEP86. As shown in Table 4, omitting the top diffuse function has a significant effect (especially for HLY, GAPT, and QTAIM) for the DZ basis set (which is simply too small), but it dwindles to very little for TZ, and is essentially nonexistent for QZ (except for a 0.004 a.u. RMSD difference for QTAIM, which probably is comparable to its numerical precision). This does suggest jun- basis sets as an avenue to make these calculations (for TZ and up) more economical and less prone to numerical problems (specifically, near-linear dependence).

maug (minimally augmented) basis sets,[105,106] in which diffuse functions are only retained for the angular momenta with occupied orbitals, go a step further still. maug-VDZ+d is equivalent to jun-VDZ +d; for maug-VTZ+d, we find that the omitted diffuse d function (related to jun-VTZ+d) causes significantly larger errors than for jun-VTZ+d, albeit still quite tolerable (0.012 a.u. for GAPT). For maug-VQZ+d, the errors caused by omitting the diffuse d and f functions are quite minimal. In all, once beyond DZ basis sets, maug appears to be quite a viable option, at least for neutral molecules or cations.

But a more relevant metric may be their performance relative to the basis set limit, for which we shall use the haV5Z+d results. As shown in the lower pane of Table 4, stripping away diffuse functions from haVDZ +d results in some degradation for HLY and especially for GAPT, which appears to be especially sensitive to diffuse functions (not quite surprising, as it is based on dipole moment derivatives). jun-VTZ+d, however, has essentially the same statistics as haVTZ+d, while for maug-VTZ+d one sees deterioration in HLY and APT. In the VQZ series, jun-VQZ+d is essentially indistinguishable in quality from the parent haVQZ, but also maug-VQZ+d represents no significant further degradation. Even QZ+d could be safely used here. Somewhat surprisingly, for APT charges, haVDZ+d is actually among the best performers.

## 3.8 | A note on the performance of CCSD for partial charge calculations

CCSD/haV5Z+d calculations with Gaussian proved an unsurmountable obstacle, and even CCSD/haVQZ+d was only possible for a

**TABLE 4** RMSD between revDSD-PBEP86 partial charges computed with jun-cc-pVnZ+d or maug-cc-pVnZ+d and the corresponding jul-cc-pVnZ+d, a.k.a., haVnZ+d, basis sets.

| | Hirshfeld | Hirshfeld-I | HLY | NBO7 | GAPT | QTAIM |
|---|---|---|---|---|---|---|
| RMSD relative to haV$n$Z+d for same $n$ | | | | | | |
| jun-cc-pV(D+d)Z | 0.003 | 0.005 | 0.041 | 0.010 | 0.032 | 0.024 |
| maug-cc-pV(D+d)Z | Equivalent to jun-cc-pV(D+d)Z | | | | | |
| cc-pV(D+d)Z | 0.006 | 0.010 | 0.034 | 0.023 | 0.026 | 0.023 |
| jun-cc-pV(T+d)Z | 0.000 | 0.001 | 0.002 | 0.001 | 0.001 | 0.005 |
| maug-cc-pV(T+d)Z | 0.001 | 0.001 | 0.009 | 0.003 | 0.012 | 0.007 |
| cc-pV(T+d)Z | 0.002 | 0.003 | 0.011 | 0.012 | 0.011 | 0.008 |
| jun-cc-pV(Q+d)Z | 0.000 | 0.000 | 0.000 | 0.001 | 0.000 | 0.003 |
| maug-cc-pV(Q+d)Z | 0.000 | 0.000 | 0.004 | 0.002 | 0.004 | 0.004 |
| cc-pV(Q+d)Z | 0.001 | 0.001 | 0.004 | 0.007 | 0.005 | 0.004 |
| RMSD from jul-cc-pV(5+d)Z for perspective | | | | | | |
| haVDZ+d | 0.004 | 0.010 | 0.016 | 0.039 | 0.004 | 0.044 |
| jun-VDZ+d | 0.005 | 0.011 | 0.039 | 0.032 | 0.032 | 0.060 |
| DZ+d | 0.007 | 0.016 | 0.030 | 0.035 | 0.026 | 0.059 |
| haVTZ+d | 0.002 | 0.003 | 0.003 | 0.016 | 0.001 | 0.020 |
| jun-cc-pV(T+d)Z | 0.002 | 0.003 | 0.004 | 0.016 | 0.002 | 0.021 |
| maug-cc-pV(T+d)Z | 0.002 | 0.004 | 0.010 | 0.016 | 0.012 | 0.022 |
| TZ+d | 0.002 | 0.005 | 0.010 | 0.021 | 0.011 | 0.021 |
| haVQZ+d | 0.000 | 0.001 | 0.001 | 0.012 | 0.000 | 0.013 |
| jun-VQZ+d | 0.000 | 0.001 | 0.001 | 0.013 | 0.001 | 0.013 |
| maug-VQZ+d | 0.000 | 0.001 | 0.004 | 0.013 | 0.004 | 0.014 |
| QZ+d | 0.001 | 0.002 | 0.004 | 0.016 | 0.005 | 0.014 |

**TABLE 5** Summary of RMSDs for partial charge calculations for closed-shell molecules of the W4-17 thermochemistry benchmark set.

| | %HF | hirshfeld | hirshfeldit | hlygat | NBO7 | GAPT | QTAIM | TAE |
|---|---|---|---|---|---|---|---|---|
| PBE | 0 | 0.016 | 0.040 | 0.019 | 0.029 | 0.036 | 0.059 | 26.987 |
| PBE10PBE | 10 | 0.011 | 0.030 | 0.015 | 0.023 | 0.030 | 0.043 | 18.242 |
| PBE20PBE | 20 | 0.008 | 0.022 | 0.013 | 0.019 | 0.028 | 0.028 | 10.473 |
| PBE0 | 25 | 0.007 | 0.020 | 0.013 | 0.018 | 0.029 | 0.021 | 7.237 |
| PBE33PBE | 33 | 0.007 | 0.018 | 0.015 | 0.018 | 0.033 | 0.015 | 5.684 |
| PBE38PBE | 38 | 0.009 | 0.019 | 0.016 | 0.019 | 0.036 | 0.017 | 7.838 |
| PBE50PBE | 50 | 0.013 | 0.026 | 0.021 | 0.024 | 0.046 | 0.030 | 15.932 |
| PBE75PBE | 75 | 0.023 | 0.046 | 0.032 | 0.039 | 0.069 | 0.066 | 33.836 |
| PBE100PBE | 100 | 0.033 | 0.066 | 0.043 | 0.054 | 0.095 | 0.102 | 50.162 |

*Note:* The heat mapping for diagnostics within each column ranges from green (indicating the lowest RMSD) to red (representing the highest RMSD). The last column corresponds to the RMSDs for the total atomization energies (TAE in kcal/mol) for the complete W4-17 dataset. Calculations are carried out using the haVTZ+d basis set, and revDSD-PBEP86 is employed as reference.

subset (albeit a sizable one), hence we only have complete CCSD data for the haVDZ+d and haVTZ+d basis sets. As our analysis in Sections 3.1–3.5 predominantly employed the haV5Z+d basis set, we opted to discuss the statistics related to CCSD separately here. We observed that across all charge types and with the haVDZ+d and haVTZ+d basis sets, MP2-based partial charges were closer to the reference (i.e., revDSD-PBEP86) than CCSD. Furthermore, CCSD was outperformed by semi-empirical double hybrids across all charge types.

## 3.9 | Effect of varying the fraction of Hartree-Fock exchange on functional performance for partial charges

As observed above, hybrid DFAs consistently exhibit superior performance compared to pure DFAs. The selection of percentage HF exchange plays a crucial role in determining the accuracy of hybrid functionals. In Table 5, we examined the performance of PBE*n* (where "n" represents the percentage HF exchange) for all charge types with the haVTZ+d basis set.

For thermochemical applications of global hybrid GGAs, exchange of about 20% (as in B3LYP[80]) or 25% (as in PBE[14]) is well-known to be optimal; if barrier heights are also given weight, one-third (33%) may be a better compromise.[107,108] In Table 5, we see that this is also the optimal range for partial charges; intriguingly, smaller percentages appear to be preferred by GAPT (20%) and HLY (20%–25%) than by iterative Hirshfeld and QTAIM (both 33%) and by NBO (25%–33%). In the same table, we also present the RMSDs pertaining to atomization energies, and we observed that our findings regarding the energetics remain consistent for partial charges too.

## 3.10 | A note on (Weigend-Ahlrichs[16]) def2 basis sets

The Weigend-Ahlrichs def2 (a.k.a., Karlsruhe def2 or Turbomole def2) basis sets are frequently employed in DFT calculations owing to their availability for almost all of the Periodic Table. We sought to determine whether our conclusions derived from Dunning basis sets (see Sections 3.1–3.6 for details) also apply to def2 basis sets, particularly the smaller ones. After analyzing def2-SVP, def2-TZVPP, and def2-QZVPP basis sets (see statistics in the Supporting Information), we found that this is indeed the case. (Note that H, B–F, and Al–Cl are *not* elements for which cc-pVnZ and def2 basis sets are identical or even just similar.)

## 4 | CONCLUSIONS

In this study, we have attempted to shed light on the sensitivity of different partial charge types to 1-particle basis set and n-particle correlation treatment (if WFT) or exchange-correlation functional (if DFT). Our findings reveal that the superiority of semi-empirical double hybrids for energetic and vibrational spectroscopic properties also extends to partial charges. In fact, semi-empirical DHDFs not only outperform the lower-rung DFAs on Jacob's Ladder but CCSD and MP2 as well. Furthermore, comparing CCSD and MP2, the latter displayed closer similarity to CCSD(T) for charges other than GAPT (which is the most sensitive to poorly described static correlation). Our study emphasized the critical role of nonlocal correlation, especially if a significant amount of nonlocal exchange is present in the underlying functional. Semilocal correlation was not essential for charge models such as Hirshfeld, Hirshfeld-I, HLY, NBO7, and QTAIM; some sensitivity to it was however seen in meta-GGAs. This study reveals that global hybrid functionals perform best for partial charges with 20%–33% Hartree-Fock exchange, like they do for energetic properties.

Hirshfeld, Hirshfeld-I, HLY, and GAPT charge population models converge rapidly with the basis set; thus, a haVTZ+d-sized basis set is adequate for reliable results. On the contrary, QTAIM and NPA exhibited slower basis set convergence, necessitating the use of a larger basis set. It is noteworthy that for both NPA and QTAIM, HF presents the slowest basis set convergence when contrasted with the correlation components of MP2 and CCSD. Moreover, it is worth mentioning that the triples term in CCSD(T), denoted CCSD(T)-CCSD, shows even faster basis set convergence.

Furthermore, our findings on the relative performance of various DFT and WFT methods remain consistent regardless of the basis set size. The same conclusions are reached for Weigend-Ahlrichs def2 basis sets as for correlation consistent basis sets.

## ACKNOWLEDGMENTS

Work on this paper was supported by the Israel Science Foundation (grant 1969/20), by the Minerva Foundation (grant 2020/05), and by a research grant from the Artificial Intelligence and Smart Materials Research Fund (in memory of Dr. Uriel Arnon), Israel. Nisha Mehta would like to acknowledge the Feinberg Graduate School for a Sir Charles Clore Postdoctoral Fellowship as well as Dean of the Faculty and Weizmann Postdoctoral Excellence fellowships. The authors would like to thank Dr. Golokesh Santra for sharing unpublished results on the performance of PBE33PBE ('one-third') for the GMTKN55 dataset, and Drs. Irena Efremenko and Margarita Shepelenko, as well as Mr. Emmanouil Semidalas, for helpful discussions.

## DATA AVAILABILITY STATEMENT

The data that support the findings of this study are openly available in Figshare at http://doi.org/10.6084/m9.figshare.24720771, reference number 24720771.

## ORCID

*Nisha Mehta*  https://orcid.org/0000-0001-7222-4108
*Jan M. L. Martin*  https://orcid.org/0000-0002-0005-5074

## SUPPORTING INFORMATION

Additional supporting information can be found online in the Supporting Information section at the end of this article.